\documentclass[sigplan, screen]{acmart}
\AtBeginDocument{%
  \providecommand\BibTeX{{%
    \normalfont B\kern-0.5em{\scshape i\kern-0.25em b}\kern-0.8em\TeX}}}

\renewcommand\footnotetextcopyrightpermission[1]{} 
\acmConference[]{}{}{}

\acmSubmissionID{123-A56-BU3}

\usepackage{array}
\usepackage{amssymb,amsmath,graphics,graphicx,latexsym,amssymb,url}
\usepackage{balance}
\usepackage{times}
\usepackage{hyperref}
\usepackage{lipsum,algorithm,multicol}

\usepackage{amssymb}
\usepackage{graphicx}
\usepackage{algpseudocode}
\usepackage{xspace,url}
\usepackage{xcolor}
\usepackage{multirow}
\usepackage{subfig}

\newcommand{\SPSE}{\textsc{KLEE\-Spectre }}

\usepackage{listings}

\definecolor{dkgreen}{rgb}{0,0.6,0}
\definecolor{gray}{rgb}{0.5,0.5,0.5}
\definecolor{mauve}{rgb}{0.58,0,0.82}
\definecolor{black}{rgb}{0,0,0}

\newtheorem{theorem}{Theorem}
\newtheorem{Property}[theorem]{Property}
\newtheorem{Definition}[theorem]{Definition}

\newcommand{\program}{\ensuremath{\mathcal{P}}}

\newcommand{\sew}{\ensuremath{\mathit{SEW}}}

\begin{document}

\title{\SPSE: Detecting Information Leakage through Speculative Cache Attacks via Symbolic Execution}

\author{Guanhua Wang}
\affiliation{%
  \institution{National University  of Singapore} 
}
\authornote{Both authors contributed equally to this research.}
\email{guanhua@comp.nus.edu.sg}
\author{Sudipta Chattopadhyay}
\authornotemark[1]
\affiliation{%
  \institution{Singapore University of Technology and Design}
}

\author{Arnab Kumar Biswas}
\affiliation{%
  \institution{National University  of Singapore}
 }
\author{Tulika Mitra}
\affiliation{%
  \institution{National University  of Singapore}
 }

\author{Abhik Roychoudhury}
\affiliation{%
  \institution{National University  of Singapore}
 }

\begin{abstract}
	Spectre attacks disclosed in the early 2018 expose data leakage scenarios via cache side channels. Specifically, speculatively executed paths due to branch mis-prediction may bring secret data into the cache which are then exposed via cache side channels even after the speculative execution is squashed. Symbolic execution is a well known test generation method to cover program paths at the level of the application software. In this paper, we extend symbolic execution with modeling of cache and speculative execution. Our tool \SPSE, built on top of the KLEE symbolic execution engine, can thus provide a testing engine to check for the data leakage through cache side channel as shown via Spectre attacks. Our symbolic cache model can verify whether the sensitive data leakage due to speculative execution can be observed by an attacker at a given program point. Our experiments show that \SPSE can effectively detect data leakage along speculatively executed paths and our cache model can further make the leakage detection much more precise.  
\end{abstract}

\maketitle

\section {Introduction}
\label{sec:introduction}
Speculative execution in modern super-scalar microprocessors improves the program performance (by reducing execution time and by increasing throughput) compared to a non-speculative processor by predicting both the outcome and the target of branching instructions. The processor continues executing instructions after the branch where the number of speculatively executed instructions depends on how soon the actual branch condition is evaluated 
and also on the size of the buffer that holds the resulting states during speculative execution.

If the prediction of a branching instruction is incorrect, all effects due to the speculatively executed instructions after the branch instruction are rolled back. To this end, the buffer and pipeline stages are flushed which hold these instructions or their results. However, if the cache content is also modified due to speculatively executed load instructions, the cache state is not fully rolled back. This opens up the possibility of a cache side channel through which an attacker can obtain sensitive information from a user who shares the same platform with the attacker. The family of Spectre attacks~\cite{Kocher2018spectre} shows that this vulnerability is present in all modern general purpose processors. Such a vulnerability thus poses 
major concerns from the stand-point of software security.

Symbolic execution~\cite{King76} is a well-known path exploration method that  can be used for program testing and verification. Given a program with un-instantiated or symbolic inputs, it constructs a symbolic execution tree by expanding both directions of every branch whose outcome depends on symbolic variable(s). The leaf nodes of the tree correspond to program paths, and by solving the constraint accumulated along a program path (also called a path condition), a test input to explore the path can be generated.

Symbolic execution can be used to cover program paths (modulo a time budget). However, it does not consider behaviors induced by performance enhancing features of the underlying processor, specifically cache and branch prediction. Due to branch mis-prediction, certain paths may be speculatively executed and then squashed. Such speculatively executed paths are not covered in symbolic execution. However, one may argue that there is no need to cover the speculatively executed paths since they are ultimately squashed and they have no impact on the observable behavior of the program. However, in the presence of caches, certain sensitive data may be brought into a cache in a speculatively executed path. This data may linger in the cache even after the speculative path is squashed. Such sensitive data may then be potentially ex-filtrated by attackers via cache side channels. Current generation symbolic execution engines, as embodied by tools like KLEE~\cite{cadar2008klee} do not demonstrate the presence or absence of such side channel scenarios. This is because the reasoning in current day symbolic execution engines is solely at the program level.

In this paper, we extend symbolic execution with the modeling of speculative execution as well as cache accesses. For an unresolved branch involving a symbolic variable, classical symbolic execution considers two possibilities - the branch is either taken or not taken. In the presence of speculative execution, note that for every unresolved branch we need to consider four possibilities, namely:  taken and correctly predicted, taken and mis-predicated, not taken and correctly predicted, not taken and mis-predicted. 
As explained earlier, since the mis-predicted execution paths are squashed, they only need to be considered in symbolic execution in the presence of cache modeling. We model the behavior of the cache by capturing memory accesses to concrete or symbolic memory addresses; the symbolic memory accesses occur when the accessed memory address depends on a symbolic input such as accessing array element {\tt a[i]} when {\tt i} is a symbolic input variable. Given such symbolic memory accesses, the possible cache conflicts (two memory accesses to the same cache set) can be captured as a symbolic formula. By solving such symbolic formula, we can enunciate whether a secret brought into cache in a speculative path continues to linger in the cache (this is when it has not been evicted from the cache due to cache conflicts). Hence we can detect and infer the cache side-channel leakage in Spectre attacks. 

The remainder of the paper is organized as follows. After providing a brief background (Section~\ref{sec:background}) and overview (Section~\ref{sec:overview}) of \SPSE, we make the following contributions: 

\begin{enumerate}
    \item We present \SPSE, our methodology to extend state-of-the-art 
    symbolic execution engines with micro-architectural features, specifically speculative execution and caches (Section~\ref{sec:methodology}). 
    
    \item We present a symbolic cache model embodied in \SPSE to precisely detect and highlight cache side-channel leakage through speculative execution paths, resulting in potential Spectre style attacks (Section~\ref{sec:methodology}). 
    
    \item We implement our \SPSE approach on top of state-of-the-art and widely used symbolic virtual machine KLEE (Section~\ref{sec:implementation}). Our implementation and all experimental data are publicly available: \\ https://github.com/winter2020/kleespectre.  
    
    \item We evaluate \SPSE on litmus tests provided by Kocher~\cite{spectremitigations} as well as on real-world cryptographic programs from {\tt libTomCrypt}, {\tt Linux-tegra}, \\ {\tt openssl} and {\tt hpn-ssh}. Our evaluation reveals that \SPSE can effectively and efficiently detect Spectre vulnerable code. Moreover, the cache modeling embodied in \SPSE results in a precise leakage detection by ruling out false positives.  
\end{enumerate}
After discussing the related work (Section~\ref{sec:related_work}), we conclude in Section~\ref{sec:conclusion}.

	
	
	

\section {Background and threat model}
\label{sec:background}
In this section, we introduce the necessary background regarding the speculative execution and our targeted threat model.  

\paragraph*{\bf{Speculative execution.}}
Speculative execution~\cite{gonzalez1997speculative} is an indispensable micro-architectural optimization for performance enhancement in modern superscalar processors. Speculative execution allows the processor pipeline to continue execution even in the presence of some data or control dependency between the current instruction and the unfinished instructions instead of stalling the pipeline. Branch predictor is one of the prediction unit in processor supporting speculative execution. The branch predictor predicts the execution path based on the history of the executed branch instructions. The processor stores a record of the speculatively executed instructions in a so-called Reorder Buffer (ROB). This buffer mainly helps the processor to commit all instructions in-order though they are executed out-of-order. If the outcome of a branch prediction is correct, then the instructions in ROB are committed to the architectural state, otherwise, the results of these instructions are squashed. However, the effect of the load execution unit i.e. the bytes that are read from memory during speculative execution may reside in the cache. The state of the cache is usually not squashed due to performance reasons. Thus, for a mis-predicted branch, even though the functional effects of all speculatively executed instructions are rolled back, the cache state may hold unexpected memory addresses. 
This phenomenon opens the potential vulnerability of cache side-channel attack.


\paragraph*{\bf{Bounds Check Bypass (BCB) attack.}}
Spectre-style attacks have proven that the computer can leak secret data through the cache side channel when it performs the speculative execution. \textbf{Bound Check Bypass} (BCB, also called Spectre variant 1) attack is one such Spectre attack. BCB attack can be performed by mis-training a vulnerable branch in the victim's process to leak data from the victim.   
\begin{lstlisting}[caption={Example code of Spectre variant 1.}\label{lst:spectev1}]
if (x < array1_size) {        \\ VB
    y = array1[x];  \\ RS
    temp |= array2[y * 64];   \\ LS
}
\end{lstlisting}
Listing~\ref{lst:spectev1} shows an example code vulnerable to BCB attacks. In this example code, 
if the condition {\tt x < array1\_size} holds, then the statement at line~2 loads {\tt array1[x]} to variable {\tt y}. Finally, the statement at line~3 reads data from {\tt array2[]} where the accessed address depends on the value {\tt array1[x]}. Normally, the boundary check at  line~1 guarantees the absence of out-of-bound memory access. 
However, in the presence of the speculative execution, such guarantees do not hold. For example, the mis-prediction of the branch instruction at line~1 allows a memory access {\tt array1[x]} where {\tt x $\ge$ array1\_size}. 
Such a memory access may point to a sensitive value. Thus, {\tt y} may hold 
a sensitive value when the branch is mis-predicted. 
Finally, the statement at line~3 changes the cache state using the potential sensitive value {\tt y}. By observing this cache state, the attacker can reconstruct the potentially sensitive value {\tt y}. 
For simplicity, we name the branch potentially causing the BCB attack as \textbf{Vulnerable Branch (VB)}, the instruction loading the potential sensitive data as \textbf{Read Secret (RS)} (e.g statement at line~2) and 
the instruction leaking the sensitive data to cache state as \textbf{Leak Secret (LS)} (e.g statement at line~3). 

	
\paragraph*{\bf{Threat model.}}
Similar to the existing literature on cache side-channel attacks~\cite{liu2015last}, in this work, we assume the 
victim and the attacker coexist on a machine, and they 
share the cache. The attacker can execute any code in 
its security domain (e.g. a process or a virtual machine) 
and it can learn information from the shared cache 
side-channel. 
Besides, in our threat model, we do not consider the data 
leakage in the normal execution path. Instead, we focus on 
data leakage only due to the speculative execution.


\begin{lstlisting}[caption={Data leakage in dead code.}\label{lst:e1}]
 int a=100, size=16; 
 char array1[16];
 char array2[256*64];
 int victim() {
	int y=0, temp=0;  
	if (a < size) {   //CB
		y = array1[a];
		temp |= array2[y];
 	}   
	return temp;
 }
\end{lstlisting}

We assume that all conditional branches in a program are potentially vulnerable. 
This is in line with the existing works on Spectre-style attacks~\cite{canella2018systematic} that show the possibility of a branch to be mis-trained either by the victim process or outside the victim process (e.g. by an attacker-controlled process). As a result, any branch in the victim process is potentially vulnerable to mis-training by the attacker. 
%
%
To consider the implication of our threat model, consider the code in listing~\ref{lst:e1}. Since the conditional branch at line~6 is 
unsatisfiable, the code at lines 7--8 will never be executed without speculation. However, in our considered threat model, the code at 
lines~7--8 can leak data if the branch at line~6 is mis-trained and 
the branch is subsequently mis-predicted (thus pointing outside the 
array bound of {\tt array1}). We also note that neither 
the branch nor the memory access at line~7 is controlled by any 
external input. 

Finally, we assume that the attacker can either perform the {\em access-based} cache side-channel attack or the {\em trace-based} cache side-channel attack~\cite{reineke2007timing}. 
The ability of such attackers depend on which execution points (s)he 
observes cache states. In particular, the access-based attack assumes 
that an attacker can probe the cache only upon the termination of a 
program. On the contrary, the trace-based attack assumes that an 
attacker can snoop the cache after any executed instruction from the 
victim process. 

\begin{figure*}[ht]
	\centering
	\includegraphics[width=1\linewidth]{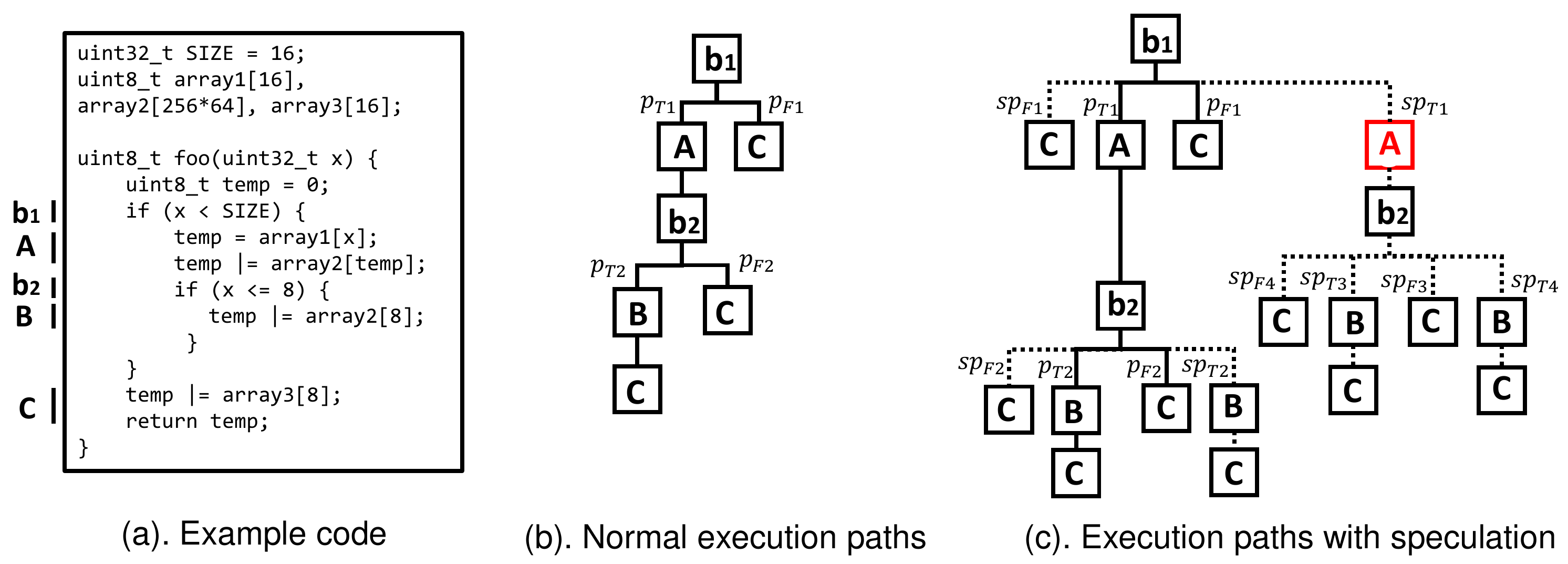}
	\caption{The example code, and its normal execution paths along with the 
	execution paths with branch speculation. (a) example code where $b_1$ and 
	$b_2$ capture branch instructions. $A$, $B$ and $C$ indicate the corresponding 
	basic blocks. (b) Execution paths explored by classic symbolic execution. 
	$p_{T\#}$, $p_{F\#}$ represent normal paths that go along the {\em true} 
	or {\em false} leg of a branch. (c) Symbolic execution tree explored by \SPSE. 
	$sp_{T\#}$, $sp_{F\#}$ denote speculative paths that go along the {\em true} or 
	{\em false} leg of a branch. The node in red color indicates the basic block 
	with potential data leakage.}
	\label{fig:example}
\end{figure*}

\section{Overview}
\label{sec:overview}
Intuitively, \SPSE is  an effort to consider and expose the 
micro-architectural execution semantics at software layer. Specifically, 
\SPSE enhances the machinery of symbolic execution with branch speculation 
and cache modeling. In the following, we will use a running example to show 
the motivation behind the design of \SPSE and briefly outline the \SPSE work-flow. 
We use the term {\em normal execution} to capture the execution semantics 
embodied in classic symbolic execution tools. 

\smallskip\noindent
\textbf{The example:} We consider the example code shown in Figure~\ref{fig:example}(a). 
The variable {\tt x} is a user controlled input. The code performs several memory 
related operations on two arrays {\tt array1} and {\tt array2}. Although {\tt x} is 
user controlled, we note that the access to {\tt array1[x]} is protected by the bound 
check (i.e. $x < \mathit{SIZE}$). Thus, considering the normal execution, the example 
does not exhibit any out-of-bound access. Figure~\ref{fig:example}(b) captures the 
execution tree generated by any classic symbolic execution tool.

\smallskip\noindent
\textbf{Enhancing symbolic execution:}
Consider the code fragments labelled {\tt A} in Figure~\ref{fig:example}(a). Such 
a code has the following problems that only appear in the presence of branch speculation. 
Assume the value of the user controlled input is such that $x \geq \mathit{SIZE}$. 
If the branch $b_1$ is mis-predicted, then the memory access {\tt array1[x]} exhibits 
an out-of-bound reference. Moreover, if {\tt array1[x]} captures a sensitive value 
(e.g. a secret), then the subsequent memory access {\tt array2[array1[x]]} 
(cf. Figure~\ref{fig:example}(a)) refers to a memory address dependent on secret value. 
Memory addresses that depend on secret values are potentially exposed to cache 
side-channel attacks. For example, consider the access-based attacker who probes the 
state of the cache after the end of execution. For such an attack, the attacker might 
be successful to ex-filtrate the value of {\tt array1[x]} (potentially holding a sensitive 
value) only if {\tt array2[array1[x]]} remains in the cache after the execution. 

It is evident from the preceding example that detecting
the potential leakage of 
{\tt array1[x]} is beyond the capability of classic symbolic execution. 
Specifically, to detect this side channel scenario, it is crucial to capture both the branch speculation and the cache behaviour while exploring symbolic execution states. 
In \SPSE, we enhance the power of symbolic execution along these two dimensions.

\smallskip\noindent
\textbf{Speculative symbolic execution in \SPSE:}
In \SPSE, the purpose of speculative symbolic execution is to {\em explore any 
potential secret that might be accessed due to branch speculation}. 
To investigate the mechanism, consider again the example in Figure~\ref{fig:example}(a). 
To incorporate the branch speculation within symbolic execution, consider the branch $b_1$ 
(i.e. $x < \mathit{SIZE}$). In the presence of branch speculation, \SPSE encounters 
the following four scenarios: 

\begin{enumerate}
	\item {$p_{T1}$:} $x < \mathit{SIZE}$ is satisfiable and the branch $b_1$ is 
	{\em correctly} predicted. In this case, the symbolic execution will fork 
	a new state with constraint $x < \mathit{SIZE}$ and proceeds by executing the 
	code fragment {\tt A}. 
	
	\item {$p_{F1}$:} $x \geq \mathit{SIZE}$ is satisfiable and the branch $b_1$ is 
	{\em correctly} predicted. In this case, the symbolic execution will fork 
	a new state with constraint $x \geq \mathit{SIZE}$ and proceeds by executing 
	the code fragment {\tt C}.  
	
	\item {$sp_{T1}$:} $x \geq \mathit{SIZE}$ is satisfiable and the branch $b_1$ 
	is {\em mis-predicted}. In this case, \SPSE forks a new state 
	with constraint $x \geq \mathit{SIZE}$, but proceeds by executing the code 
	fragment {\tt A}. 
	
	\item {$sp_{F1}$:} $x < \mathit{SIZE}$ is satisfiable and the branch $b_1$ is 
	{\em mis-predicted}. \SPSE forks a new state 
	with constraint $x < \mathit{SIZE}$, but proceeds by executing the code fragment 
	{\tt C}. 
	
\end{enumerate}
$sp_{T1}$ and $sp_{F1}$ are the additional symbolic states explored by \SPSE at branch 
$b_1$. Figure~\ref{fig:example}(b) and (c) capture the symbolic execution trees explored 
by normal symbolic execution and \SPSE, respectively, for the code in Figure~\ref{fig:example}(a). 

The symbolic execution along a speculative path spans across only a limited number 
of instructions. This is because the maximum number of speculatively executed instructions is bounded
by the size of the re-order buffer (ROB). In \SPSE, we use Speculative Execution Window 
(\sew) to limit the number of speculatively executed instructions at any branch. It is 
worthwhile to note that a speculatively executed path may still span over multiple branch 
instructions (cf. Figure~\ref{fig:example}(c)) despite the limited size of \sew.

\SPSE prunes speculative symbolic states if they do not pose any risk of data leakage. 
For example, in Figure~\ref{fig:example}(c), only the execution of code fragment {\tt A} 
under the branch $sp_{T1}$ exhibits such risk. 
This is due to the access of array elements {\tt array1[x]} and {\tt array2[array1[x]]}. Also note that, the symbolic states 
$sp_{T3}$, $sp_{F3}$, $sp_{T4}$ and $sp_{F4}$ are all discarded once \SPSE reaches 
the limit of speculation window \sew. In this fashion, \SPSE can control the explosion 
in the number of symbolic states due to speculation. Specifically, for the example 
in Figure~\ref{fig:example}(c), \SPSE only keeps the record of executing code {\tt A} 
under the speculative state $sp_{T1}$. 

The next stage of \SPSE computes whether the secret accessed in $sp_{T1}$ can 
potentially be ex-filtrated by a cache side-channel attacker.

 
\begin{figure}[ht]
	\centering
	\includegraphics[width=0.95\linewidth]{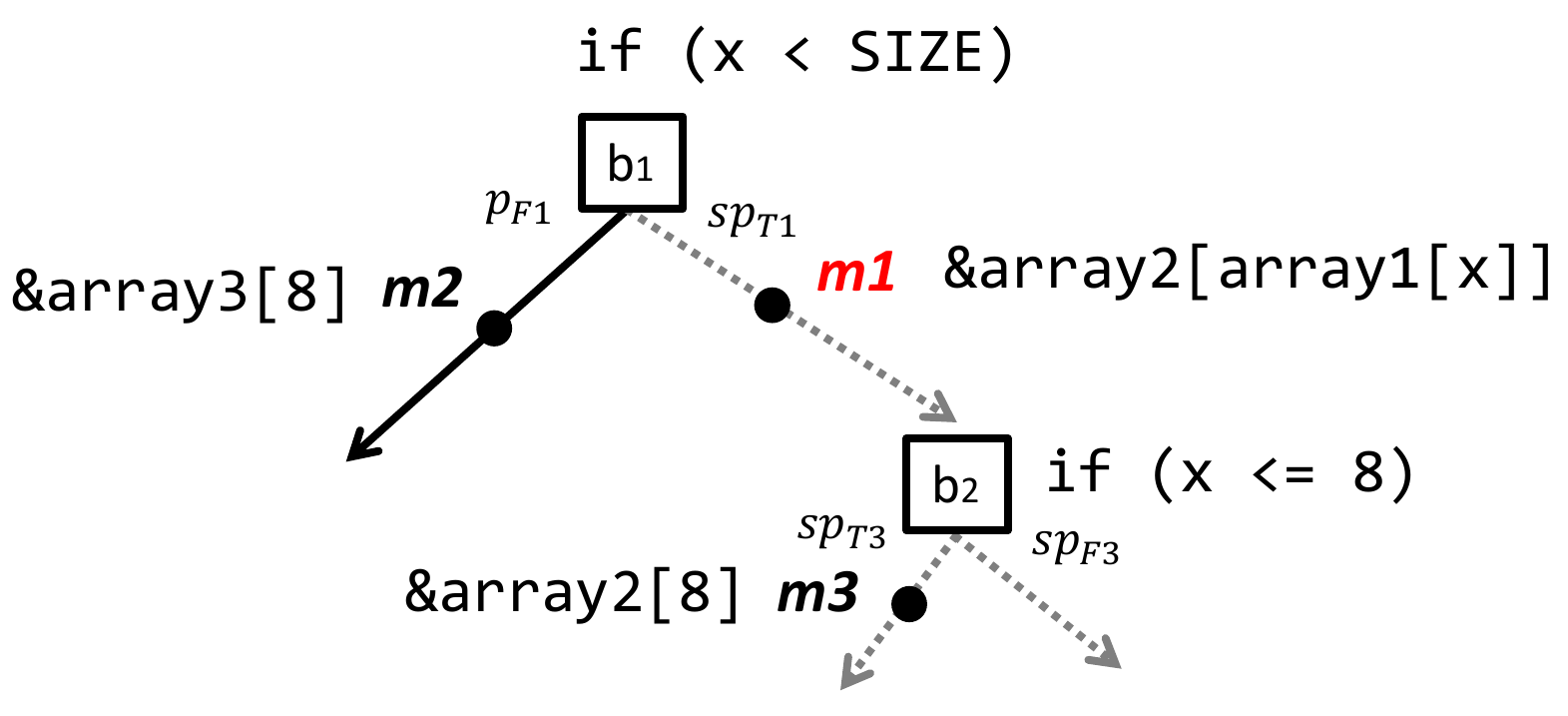}
	\caption{Partial speculative execution paths of example code. $m_{\#}$ represents a memory access on a path. The memory access in red color brings in a sensitive cache state.}
	\label{fig:example_cache}
\end{figure}

\smallskip\noindent
\textbf{Cache modeling in \SPSE:}
\SPSE computes the set of memory 
access sequences that are potentially vulnerable to a cache side-channel 
attack. Each such memory access sequence may involve at least one memory 
access along the speculative path and multiple memory accesses along the 
normal execution path. Moreover, along the speculative path, we only record memory accesses that are dependent on secret. This is because \SPSE focuses to discover data leakage due to branch speculation. 

For example, in Figure~\ref{fig:example}(c), \SPSE computes the following sequence of memory accesses for inspecting the leakage of data: 
{\small
\begin{equation*}
\langle \langle A, x > \mathit{SIZE}, \&array2[array1[x]] \rangle, \langle C, x > \mathit{SIZE}, \&array3[8] \rangle \rangle 
\end{equation*}}
The triplet $\langle A, x > \mathit{SIZE}, \&array2[array1[x]] \rangle$ captures that the memory address $\&array2[array1[x]]$ was accessed with the symbolic constraint $x > \mathit{SIZE}$ in code fragment {\tt A}. The sequence of memory accesses capture the accesses in the speculative state $sp_{T1}$ followed by 
a memory access in the normal state $p_{F1}$ (cf. Figure~\ref{fig:example}(c)).  
Even though the functional states in $sp_{T1}$  do not affect the computation 
in $p_{F1}$, the cache state influenced in $sp_{T1}$ remains unchanged when  the branch is resolved and the execution continues through code fragment 
{\tt C} (cf. Figure~\ref{fig:example_cache}). 

Through our cache modeling, we check the presence of the address $\&array2[array1[x]]$ 
in the cache when the code segment {\tt C} finishes execution. To this end, we check 
whether memory access $array3[8]$ can replace $array2[array1[x]]$ from the cache. For 
the sake of simplicity, let us assume a 1-way associative (i.e. direct-mapped) 
cache. For direct-mapped caches, a memory address maps to exactly one cache line. 
In particular, the following symbolic condition is satisfiable if and only if the 
terminating cache state holds the memory address $\&array2[array1[x]]$: 
\begin{equation}
\label{eq:example-cache}
\begin{split}
\left ( x > \mathit{SIZE} \right ) \wedge 
\left ( set \left ( \&array2[array1[x]] \right ) \ne set \left ( \&array3[8] \right ) \right ) 
\\
\vee \left ( tag \left ( \&array2[array1[x]] \right ) = tag \left ( \&array3[8] \right ) \right )
\end{split}
\end{equation}
where $set(x)$ and $tag(x)$ capture the cache line and cache tag, respectively, 
for a memory address $x$. Intuitively, the constraints in Formula~\ref{eq:example-cache} 
can be presented to a satisfiability modulo theory (SMT) solver. \SPSE 
formulates such constraints for each memory access sequence that may access secrets 
along a speculative path. These constraints are then discharged by an SMT solver to 
check the presence of data leakage due to speculation. 

In the subsequent section, we will elaborate the individual sub-systems within 
\SPSE in detail.

\section{Cache Aware Speculative Symbolic Execution}
\label{sec:methodology}
In this section, we describe the design of \SPSE. First, 
we describe the overall speculative symbolic execution 
process augmented with a symbolic cache model. Subsequently, 
we discuss in detail the features of the cache model to 
accurately detect the cache side-channel leakage along 
a speculative execution path.

\subsection{Speculative Symbolic Execution}
Algorithm~\ref{alg:symexec} outlines the overall process involved in \SPSE. 
Our methodology takes a program \program\ and symbolically executes \program\  
by taking into account the speculation at branches. Moreover, \SPSE records 
memory accesses along the speculatively executed paths to check whether any 
such memory access may refer to a secret. Finally, the sequence of memory accesses are 
used to formulate a symbolic cache model $\Gamma_{spectre}$. The model  
$\Gamma_{spectre}$ is satisfiable if a possible secret $s$, accessed along 
a speculative path, remains in the cache after the program execution.  This is because 
the presence of a speculatively accessed secret $s$ in the cache might result 
in ex-filtrating $s$ via a cache side-channel attack. 
The construction of the speculative execution 
revolves around the concept 
of speculative execution window ($\sew$). Such a bounded window 
captures the number of instructions that a processor might speculatively 
execute beyond a branch before the outcome of the branch is resolved. 
We note that $\sew$ may span across multiple unresolved branch 
instructions. 
 

At a broader perspective, Algorithm~\ref{alg:symexec} intercepts each conditional 
branch instruction $r$ and explores all possible speculatively executed instructions 
from this branch. To this end, we compute $\Omega_r$. After handling a conditional branch $r$, each element in $\Omega_r$ is 
a possible sequence of memory accesses that might have occurred during a 
speculative execution from $r$. Moreover, $\Omega_r$ only records memory accesses that may refer to a secret. If memory accesses do not refer to secrets along speculatively executed paths, then they do not impose any risk related to the leakage of information. Algorithm~\ref{alg:symexec} terminates when the time budget exceeds or \SPSE explores all (speculatively) executed paths and $\Omega_r$ is 
constructed for every conditional branch $r$. In the following, we will discuss some critical features of \SPSE. 


\paragraph*{\bf Identifying Secret Access:} In this work, we identify secrets 
as follows. For each memory-related instruction 
$r$, we consider that $r$ accesses a secret if and only if $r$ points to an 
out-of-bound memory location. Although all such memory accesses may not 
refer to secrets, these memory accesses capture illegal accesses, a typical 
target for attacks exploiting speculative execution. Nevertheless, \SPSE can 
easily be configured for explicitly marked secret data, such as a secret key 
in an encryption routine. We use the function $\mathit{DEP}(sec,m)$ to capture 
whether some memory address $m$ is data-dependent on secret $sec$. Concretely, 
$\mathit{DEP}(sec,m)$ is {\em true} if and only if $m$ is data-dependent on the secret $sec$. 

%
%
%

\paragraph*{\bf Procedure {\tt SPSE}} 
Algorithm~\ref{alg:symexec} outlines the symbolic execution process embodied in 
\SPSE. Intuitively, \SPSE modifies the handling of branch instructions within 
a classic symbolic execution process. For each conditional branch $r$, \SPSE 
maintains a structure $\Omega[r]$. Upon encountering a conditional branch 
instruction $r$, \SPSE explores all possible execution paths that might occur 
due to branch speculation. This is accomplished via the procedure 
{\tt ExpandSpecTree}. Consider the symbolic state before the branch instruction 
is $\mu$ and the branch condition is $\phi_r$. If $\pi[\mu]$ captures the 
partial path condition before $r$, then a speculative execution may proceed 
in the following two scenarios. Firstly, the true leg of the branch might be 
explored with the constraint $ \pi[\mu] \wedge \neg \phi_r$. Secondly, the 
false leg of the branch might be explored with the constraint 
$ \pi[\mu] \wedge \phi_r$. These explorations are accomplished via the two 
calls to procedure {\tt ExpandSpecTree} in Algorithm~\ref{alg:symexec}. 
Upon termination of {\tt ExpandSpecTree} for a branch instruction $r$, the 
structure $\Omega[r]$ contains the set of memory access sequences that 
depend on some secret. Therefore, these memory accesses are candidates that 
may leak secret information via cache side channel. Each memory access 
captures a triplet of the form $\langle r, \pi, \sigma \rangle$ where 
$r$ points to the instruction in the execution trace, $\pi$ captures 
the symbolic constraint with which $r$ was executed and $\sigma$ captures 
the symbolic expression of the accessed memory address. Finally, 
\SPSE records all memory accesses that influence the cache state for 
memory blocks in $\Omega[r]$. Thus, after termination of a symbolically executed path, each list $\Gamma \in \Omega[r]$ contains all memory accesses that may replace a memory block accessed during the speculation at $r$.


\paragraph*{\bf Procedure {\tt ExpandSpecTree}:} 
Algorithm~\ref{alg:symexec-tree} outlines the overall process of exploring the set of 
speculative execution paths. In summary, {\tt ExpandSpecTree} performs the following 
operations. First, it explores all speculative paths until the speculation depth 
\sew. We note that such an exploration may involve nested speculation. Secondly, while 
exploring the speculative paths, we record memory addresses for checking information 
leakage through the cache. These are the set of memory accesses that may depend on 
some secret $sec \in \mathit{SEC}$. In our framework, we consider that any out-of-bound 
memory access along a speculative path points to a secret. Thus, the procedure 
{\tt ExpandSpecTree} also records the potential secrets during exploration.

\paragraph*{\bf Termination of a speculative state. } The execution of speculative state 
can be terminated in the following ways:

\begin{enumerate}
	\item The speculation window {\sew} expires. Since {\sew} captures the maximum number 
	of instructions that can be executed speculatively, we terminate the exploration of 
	a speculative execution state after exploring {\sew} instructions. 
	\item A memory fence instruction is executed. The memory fence can stop the speculative 
	execution triggered due to branch mis-prediction. 
	\item An exception is raised. When an exception (e.g. divide by zero) is raised, the 
	speculative execution terminates. This is analogous to the termination of normal 
	execution.     
\end{enumerate}

Algorithm~\ref{alg:symexec} satisfies the following crucial properties: 

\begin{Property}
	Consider an instance of the procedure call \\ {\tt ExpandSpecTree}($\pi$, $\mu$, $r$, $r_s$, $\Gamma$). 
	Upon termination of this call, let us assume $\Omega[r] = \{\Gamma_1, \Gamma_2, \ldots, \Gamma_n\}$. 
	During an arbitrary execution, further assume that the conditional branch 
	$r$ was mispredicted and memory address $m_s$ was accessed speculatively. 
	If $m_s$ is data-dependent on some secret, then  
	$\langle *, *, m_s \rangle \in \Gamma_i$ for some $\Gamma_i \in \Omega[r]$. 
	In short, $\Omega[r]$ is guaranteed to be an over-approximation of speculatively 
	accessed memory addresses that are dependent on secret. 
\end{Property}

\begin{Property}
    Consider $\Gamma \in \Omega[r]$ after the termination of a symbolically executed path with the symbolic state $\mu$. 
    Let  $\langle *, *, m_s \rangle \in \Gamma$ where $m_s$ is data-dependent on some 
    secret. Assume $tail \left (\Gamma, \langle *, *, m_s \rangle \right )$ captures 
    the set of elements in the sequence $\Gamma$ post the element $\langle *, *, m_s \rangle$. 
    If $\langle *, *, m \rangle \in tail \left (\Gamma, \langle *, *, m_s \rangle \right )$, 
    then the memory block $m$ must be accessed following the access to $m_s$ for any concrete 
    execution realizing the symbolic state $\mu$. 
\end{Property}



\begin{algorithm}[t]
	\caption{Symbolic execution process embodied in \SPSE}
	\label{alg:symexec}
		{\small
			\begin{algorithmic}
			\Procedure{\SPSE}{\program, \sew}
				\State{Let $r$ be the first instruction in \program}
				\State{\textsf{/* $\mu_{0}$ is the initial state before running \program\ */}}
				\State{\textsf{/* $\pi[\mu_{0}]$ is the constraint associated with the state $\mu_0$ */}}
				\State{$\chi$ := $\{\mu_{0}\}$; $\pi[\mu_{0}]$ := {\em true}; $Spec$ := $\emptyset$}
				\While{$\chi \ne \emptyset$}
				\State{Choose a symbolic state $\mu \in \chi$}
				\State{\textsf{/* get the next instruction to symbolically execute */}}
				\State{$r$ := {\tt GetNextInstruction}(\program, $\mu$)}
				\State{$\mu$ := {\tt ExecuteSymbolic}($\mu$, $r$)}
				\If{$r$ is a conditional branch}
				\State{ Let $\phi_{r}$ be the branch condition}
				\State{ remove $\mu$ from $\chi$} 
				\State{ $\Gamma_{t}$ := $\Gamma_{f}$ := empty;  $\Omega[r]$ := $\emptyset$}
				\State{$\mu_t$ := $\mu_f$ := $\mu$}
				\State{$\pi[\mu_{t}]$ := $\pi[\mu] \wedge \phi_r$;\ \ \ $\pi[\mu_{f}]$ := $\pi[\mu] \wedge \neg \phi_r$} 
				\If{$\pi[\mu_{t}]$ is satisfiable}
					\State{$r_n^t$ := {\tt GetNextInstruction}(\program, $\mu_t$)}
					\State{\textsf{/* $r_n^f$ is executed when $r$ is mis-predicted */}}
					\State{{\tt ExpandSpecTree}($\pi[\mu_{t}]$, $\mu_t$, $r$, $r_n^f$, $\Gamma_{f}$)}
					\State{$\chi$ := $\chi \cup \{\mu_t\}$}
				\EndIf
				\If{$\pi[\mu_{f}]$ is satisfiable}
					\State{$r_n^f$ := {\tt GetNextInstruction}(\program, $\mu_f$)}
					\State{\textsf{/* $r_n^t$ is executed when $r$ is mis-predicted */}}
					\State{{\tt ExpandSpecTree}($\pi[\mu_{f}]$, $\mu_f$, $r$, $r_n^t$, $\Gamma_{t}$)}
					\State{$\chi$ := $\chi \cup \{\mu_f\}$}
				\EndIf
%
				\If {$\Omega[r] \ne \emptyset$}
					\State{$Spec$ := $Spec \cup \{r\}$} 
				\EndIf
				\EndIf
				\State{\textsf{/* record memory accesses along the normal path */}}
				\If{$r$ is a memory-related instruction}					
					\State{Let $\sigma$ be the accessed memory address}
					\For {each $i \in Spec\ s.t.\  \pi[\mu] \implies \phi_i \wedge \Gamma \in \Omega[i]$}
							\State{{\tt Append} $\left ( \Gamma, \langle r, \pi[\mu], \sigma \rangle \right ) $}
					\EndFor  
				\EndIf
				\EndWhile  
			\EndProcedure
		\end{algorithmic}
		}
\end{algorithm}

\begin{algorithm}[t]
    \caption{Exploring speculative execution paths for branch $r$}
	\label{alg:symexec-tree}		
	{\small
	    \begin{algorithmic}						
				\Procedure{ExpandSpecTree}{$\pi$, $\mu$, $r$, $r_s$, $\Gamma$}
				\State{$\mu'$ := {\tt ExecuteSymbolic} ($\mu$, $r_{s}$)}
				\While{$\Delta(r, r_{s}) \leq \sew \wedge r_{s} \ne$ exit}
				\If{$r_{s}$ is a conditional branch}
				\State{Let $\phi_{rs}$ be the branch condition for $r_s$}
				\State{Let ${r}_s^t$ immediately follows if $r_s$ is taken}
				\State{Let ${r}_s^f$ immediately follows if $r_s$ is not taken}
				\If{$\pi \wedge \phi_{rs}$ is satisfiable}
					\State{$\Gamma_t^t$ := $\Gamma_t^f$ := $\Gamma$}	
					\State{\textsf{/* explore the true leg for correct prediction */}}	
					\State{{\tt ExpandSpecTree}($\pi \wedge \phi_{rs}$, $\mu'$, $r_s$, $r_s^{t}$, $\Gamma_t^t$)}
					\State{\textsf{/* explore the false leg for mis-prediction */}}
					\State{{\tt ExpandSpecTree}($\pi \wedge \phi_{rs}$, $\mu'$, $r_s$, $r_s^{f}$, $\Gamma_t^f$)}
				\EndIf 
				\If{$\pi \wedge \neg \phi_{rs}$ is satisfiable}	
					\State{$\Gamma_f^t$ := $\Gamma_f^f$ := $\Gamma$}		
					\State{\textsf{/* explore the false leg for correct prediction */}}
					\State{{\tt ExpandSpecTree}($\pi \wedge \neg \phi_{rs}$, $\mu'$, $r_s$, $r_s^{f}$, $\Gamma_f^f$)}
					\State{\textsf{/* explore the true leg for mis-prediction */}}
					\State{{\tt ExpandSpecTree}($\pi \wedge \neg \phi_{rs}$, $\mu'$, $r_s$, $r_s^{t}$, $\Gamma_f^t$)}

				\EndIf 
				\EndIf
				\If{$r_{s}$ is a memory-related instruction}
				\State{Let $\sigma_s$ be the accessed memory address}
				\State{\textsf{/* record memory access dependent on secret */}}
				\If{$\exists \mathit{sec} \in \mathit{SEC}$ such that $\mathit{DEP}(\mathit{sec},\sigma_s)$}
					\State{{\tt Append} $\left ( \Gamma, \langle r_s, \pi, \sigma_s \rangle \right ) $}
				\EndIf
				\If {$\sigma_s$ refers to a potential secret}
					\State{\textsf{/* $val(\sigma_s)$ captures value at location $\sigma_s$ */}}
					\State{$\mathit{SEC}$ := $\mathit{SEC} \cup \{val(\sigma_s) \}$ }
				\EndIf
				\EndIf
				\State{$r_{s}$ := {\tt GetNextInstruction}($\program$, $r_{s}$)}
				\EndWhile
				\State{$\Omega[r]$ := $\Omega[r] \cup \{\Gamma\}$}
				\EndProcedure		
		\end{algorithmic}						
		
		}
\end{algorithm}

\subsection{Symbolic Model of Cache}
\label{sec:cache}

In this section, we will model the cache behaviour of an execution 
path to check whether a secret remains in the cache after program 
execution. Note that our modified symbolic execution already takes 
into account the speculative execution semantics. Thus, the obtained 
execution path already accounts for memory references accessed 
speculatively. Concretely, the input to our cache model is any 
memory access sequence $\Gamma \in \Omega[r]$ (see Algorithm~\ref{alg:symexec}) 
where $\Omega[r]$ is constructed for every conditional branch 
instruction $r$. In the following, we show the cache modeling 
for a memory access sequence $\Gamma$. Since $\Gamma$ is arbitrary, 
the same modeling principle is employed for all the memory access 
sequences recorded. Concretely, any memory access sequence $\Gamma$ 
is captured by a sequence of triplets as follows: 
\begin{equation}
\label{eq:trace}
\Gamma \equiv \langle (r_1, \pi_1, \sigma_1), (r_2, \pi_2, \sigma_2), \ldots, (r_N, \pi_N, \sigma_N) \rangle
\end{equation}
where $r_i$ is a memory-related instruction, $\pi_i$ is the symbolic 
constraint with which $r_i$ was executed and $\sigma_i$ is the memory 
address accessed by $r_i$. We note that $r_i$ can be accessed along 
a speculative path or a normal path (cf. Algorithm~\ref{alg:symexec}). 
Before discussing the cache model, we  first explain the basic design principle behind caches. 

\paragraph*{\bf Basics of Cache Design:} Caches are fast memory employed 
between the CPU and the main memory. While accessing a memory location, 
the CPU first checks whether the memory location is cached. If the location 
is cached, then the CPU fetches the respective value from the cache. 
Otherwise, it accesses main memory and updates the cache with the accessed 
memory location and its value. The design parameters of a cache can be 
captured via a triplet: $\langle 2^{S}, 2^{B}, \mathcal{A} \rangle$. 
$2^{S}$ captures the number of {\em cache sets} and $2^{B}$ captures the 
size of cache line (in bytes). Each cache set can hold $\mathcal{A}$ cache 
lines while $\mathcal{A}$ is called the {\em associativity} of the cache. 
For any memory-related instruction $r$, let us assume it accesses the memory 
address $m$. The address $m$ is mapped to the cache set 
$\left ( \left \lfloor \frac{m}{2^{B}} \right \rfloor \ \mod\ 2^{S} \right )$. 
Since multiple memory addresses can be mapped to the same cache set, each cache 
line in a cache set stores a {\em tag}. This {\em tag} is identified 
via the most-significant $B$ bits of the memory address $m$. 
Once a cache set is full (i.e. holds $\mathcal{A}$ cache lines) and a new 
memory location is mapped to the same cache set, then a replacement policy 
is employed to evict a cache line and make room for fresh memory locations. 
In this work, we instantiate \SPSE for the {\em least recently used} (LRU) 
replacement policy. In LRU, the least recently accessed memory location 
in a cache set is chosen for eviction to accommodate fresh memory blocks. 
We define a {\em cache set state} as an ordered $\mathcal{A}$-tuple where 
the rightmost element captures the least recently used cache line. For 
example, in a two-way associative cache, the state $\langle L_1, L_2 \rangle$ 
captures that $L_2$ (respectively, $L_1$) is the least (respectively, most) 
recently used cache line. 

In line with the preceding description of cache design, we will assume the 
following notations throughout the section: 
\begin{itemize}
	\item $2^{S}:$ The number of cache sets in the cache. 
	\item $2^{B}:$ Size of the cache line. 
	\item $N_{s}:$ The set of memory-related instructions accessing symbolic memory addresses (i.e. potential secrets accessed along speculative paths) in memory access sequence $\Gamma$ (cf. Equation~\ref{eq:trace}).  
	\item $N_{t}:$ The set of memory-related instructions exhibited along normal path in memory access sequence $\Gamma$ (cf. Equation~\ref{eq:trace}). We note that $|N_s \cup N_t| = N$ holds. 
	\item $\mathcal{A}:$ Associativity of the cache.  
	\item $\sigma_i:$ Memory address accessed by instruction $r_i$. 
	\item $set(r_i):$ Cache set accessed by memory-related instruction $r_i$. 
	\item $tag(r_i):$ Cache tag related to the memory-related instruction $r_i$. 
\end{itemize}

It is worthwhile to note that the {\em symbolic address} is defined 
to be a memory address that is dependent on a secret {\em value}. Moreover, as mentioned in the preceding section, we only consider secrets that might be accessed along speculative paths. 
Thus, if any address dependent on secrets remains in the cache after 
program execution, the respective program is vulnerable to Spectre 
attacks. The set of instructions accessing such symbolic addresses, 
i.e., $N_s$ were identified during our novel symbolic execution stage. 

\paragraph*{\bf Cache Conflict:} The symbolic model of the cache revolves around the notion 
of {\em cache conflict}. Intuitively, the phenomenon of cache conflict influences the states 
of each cache set. This, in turn, decides whether a value is cached during or after the 
execution. In the following, we first formally define the notion of cache conflict. 

\begin{Definition}
\label{def:ccf}
\textbf{(Cache Conflict):} Consider memory-related instructions $r_i$ and $r_j$. 
Let $\zeta_i$ (respectively, $\zeta_j$) be the cache state immediately after $r_i$ 
(respectively, $r_j$) is executed. $r_j$ generates 
a cache conflict to $r_i$ only if $r_j$ is executed after $r_i$ and executing $r_j$ can influence the relative position 
of memory block $\left \lfloor \frac{\sigma_i}{2^{\mathcal{B}}} \right \rfloor$
within the cache state $\zeta_j$.
\end{Definition}

The preceding definition of cache conflict works for arbitrary memory-related instructions $r_i$ and $r_j$. In \SPSE, however, 
our objective is to check whether any symbolic address remains 
in the cache. To this end, we only need to capture the cache 
conflict when $r_j \in N_t$ and $r_i \in N_s$. The cache conflicts 
within normal paths and within the speculative paths are ignored. 
Similarly, we do not need to check whether a memory block accessed 
in normal path can be replaced via a memory block accessed along 
speculative paths. Thus, we can ignore the cache conflict when 
$r_j \in N_s$ and $r_i \in N_t$. We formalize the aforementioned 
notion of cache conflict in \SPSE via the following definition: 
\begin{Definition}
\label{def:ccf-spse}
\textbf{(Cache Conflict in \SPSE):} Consider memory-related instructions $r_i$ and $r_j$. For \SPSE, we consider a cache conflict from $r_j$  to $r_i$ if and only if $r_j$ generates a cache conflict to $r_i$ according 
to Definition~\ref{def:ccf} and $r_j \in N_t$ and $r_i \in N_s$.
\end{Definition}

By considering the notion of cache conflict, as defined in 
Definition~\ref{def:ccf-spse}, we greatly simplify the size 
of the symbolic cache model and keep the overall complexity 
of \SPSE under check. 
In the next sections, we shall elaborate the crucial conditions 
required for the generation of cache conflicts and usage of 
such conditions to check the residency of a memory block in the 
cache. Subsequently, we build upon such conditions to formulate 
the symbolic model for identifying Spectre vulnerabilities.

\paragraph*{\bf Cache Set and Cache Tag}: We note that due to the symbolic 
memory addresses, $set(r_i)$ and $tag(r_i)$ can be symbolic expressions. 
Specifically, $set(r_i)$ and $tag(r_i)$ are computed as follows: 

\begin{equation}
\label{eq:set-symbolic}
set(r_i) = (\sigma_i \gg B)\ \&\ (2^S - 1)\ subject\ to\ \pi_i
\end{equation}
\begin{equation}
\label{eq:tag-symbolic}
tag(r_i) = \sigma_i \gg (B+S) \ subject\ to\ \pi_i
\end{equation}

\paragraph*{\bf Cache Conflict and Conflict Propagation}: Our objective 
is to discover whether any 
symbolic memory address can be evicted from the cache after being accessed. 
As stated in Definition~\ref{def:ccf-spse}, \SPSE only considers cache 
conflict from memory accesses along normal path (i.e. set $N_t$) 
to the memory accesses along speculative paths (i.e. set $N_s$). 
However, it is not sufficient to check the cache conflict from 
$r_j$ ($\in N_t$) to $r_i$ ($\in N_s$) to precisely identify Spectre vulnerabilities. To check whether the conflict actually influences the relative position 
of the memory block till the end of the execution, we need to check 
whether the memory block accessed by $r_i$ can be reloaded after $r_j$ 
and before the end of the execution. If $r_i$ is reloaded after $r_j$, 
then the cache conflict generated by $r_j$ is not propagated until 
the end of the execution. Finally, we need to check whether the memory 
block accessed by $r_i$ is replaced from the cache before the execution 
terminates. This is accomplished by checking whether the number of unique cache conflicts to $r_i$ that propagate till the end of execution exceeds the cache associativity ($\mathcal{A}$). In the following, we will model these phenomenon symbolically. 

If $r_j$ generates a cache conflict to $r_i$, then the following condition
must hold: 
$r_j$ and $r_i$ access the same cache set, but have different memory-block tags. 
This is formalized as follows: 
\begin{equation}
\label{eq:conflict-1}
\psi_{cnf} \left ( r_i, r_j \right ) \equiv \left ( set(r_i) = set(r_j) \right ) 
\wedge \left ( tag(r_i) \neq tag(r_j) \right )
\end{equation}

Additionally, we need to check whether $r_j$ is a unique cache conflict. To this 
end, we check that none of the memory accesses after $r_j$ accesses the same 
memory block as $r_j$. Thus, we only account for the last memory-related 
instruction accessing the block $\left \lfloor \frac{\sigma_j}{2^B} \right 
\rfloor$. This is formalized as follows: 
\begin{equation}
\label{eq:unique}
\begin{split}
   \psi_{unq}
   \left  ( r_j \right ) 
   \equiv 
   \\
   \bigwedge_{k \in (j, N] \wedge r_k \in N_t} \left ( set(r_j) \neq set(r_k) \right ) 
\vee \left ( tag(r_j) \neq tag(r_k) \right )
\end{split}
\end{equation}

Finally, we need to check that $r_i$ is not reloaded after $r_j$. Otherwise, the 
memory block accessed by $r_i$ will be reloaded to the cache and the conflict due 
to $r_j$ would be nullified. This is formalized as follows: 
\begin{equation}
\label{eq:reload}
\psi_{rel} \left  ( r_i, r_j \right ) \equiv \bigwedge_{k \in (j, N]} \left ( set(r_i) \neq set(r_k) \right ) 
\vee \left ( tag(r_i) \neq tag(r_k) \right )
\end{equation}

Combining Equations~\ref{eq:conflict-1}-\ref{eq:reload}, we can obtain the symbolic 
condition where $r_j$ changes the relative position of the memory block accessed by
$r_i$ and such a change in the relative position of the memory block is also 
propagated until the end of the execution. Thus, when all the conditions 
$\psi_{cnf} \left ( r_i, r_j \right )$,  $\psi_{unq} \left  ( r_j \right ) $ and 
$\psi_{rel} \left  ( r_i, r_j \right )$ hold, we can say that the conflict generated 
by $r_j$ to $r_i$ is propagated until the end of the execution. This is symbolically 
captured as follows: 

\begin{equation}
\label{eq:positive-conflict}
\Theta_{j,i}^{+} \equiv \psi_{cnf} \left ( r_i, r_j \right ) \wedge \psi_{unq} \left  ( r_j \right ) 
\wedge \psi_{rel} \left  ( r_i, r_j \right ) \Rightarrow \left (cnf_{i,j} = 1 \right ) 
\end{equation}
\begin{equation}
\label{eq:negative-conflict}
\Theta_{j,i}^{-} \equiv \neg \psi_{cnf} \left ( r_i, r_j \right ) \vee \neg \psi_{unq} \left  ( r_j \right ) 
\vee \neg \psi_{rel} \left  ( r_i, r_j \right ) \Rightarrow \left (cnf_{i,j} = 0 \right ) 
\end{equation}

\paragraph*{\bf Attack Identification}:
We note that $r_j$ is arbitrary in the preceding discussion. To check whether the 
memory block accessed by $r_i$ can be replaced, we need to repeat the computation 
of $\Theta_{j,i}^{+}$ and $\Theta_{j,i}^{-}$ for any $j \in [i+1, N]$ where $N$ is 
the total number of memory accesses in the trace. Finally, we need to check whether 
the collective sum of $cnf_{i,j}$ for $j \in [i+1, N]$ exceeds the cache 
associativity. Let us assume that $spec_i$ is true if and only if the memory block 
accessed by $r_i$ may remain in the cache after program execution, thus exhibiting 
a potential Spectre attack. The truth value of $spec_i$ can be symbolically computed 
as follows: 
\begin{equation}
\label{eq:positive-spectre}
\lambda_{i} \equiv \left ( \sum_{j \in [i+1,N] \wedge r_j \in N_t}  cnf_{i,j} < \mathcal{A} \right ) \Rightarrow spec_i
\end{equation}

\paragraph*{\bf Putting it altogether}:
Finally, spectre attacks can be targeted for any memory-related instruction accessing 
a symbolic address. Therefore, Equations~\ref{eq:conflict-1}-\ref{eq:positive-spectre} need 
to account for all such symbolic memory accesses. Recall that $N_s$ captures the set 
of all memory-related instructions in the trace that access symbolic memory address. 
Thus, to check the possibility of Spectre attacks for an arbitrary (combination) of 
memory addresses, the following symbolic model is used: 
\begin{equation}
\label{eq:all-spectre}
\boxed{
\begin{aligned}
	\Gamma_{spectre} \equiv
	\\ 
	\bigwedge_{r_i \in N_s} \left ( \lambda_{i} \wedge 
	\left ( \bigwedge_{j \in [i+1,N] \wedge r_j \in N_t}  \Theta_{j,i}^{+} \wedge \Theta_{j,i}^{-} \right ) \right ) \wedge \left ( \bigvee_{r_i \in N_s} spec_i \right ) 
\end{aligned}
}
\end{equation}

We note that $\Gamma_{spectre}$ is true if and only if any of the 
symbolic memory address remains in the cache after program execution, 
thus leading to a potential spectre attack. 


\section {Implementation}
\label{sec:implementation}
\SPSE is primarily implemented on top of the state-of-the-art symbolic execution engine KLEE v2.0~\cite{cadar2008klee}. \SPSE is built from 
CLang v6.0 and it takes the LLVM bitcode generated with LLVM 6.0 as 
input. If a subject program contains external function calls, then the program is linked with KLEE- uClibc~\cite{uclib} first, before being passed to \SPSE. We used the SMT solver STP~\cite{ganesh2007decision} to check the satisfiability of the path constrains and the symbolic cache model. 
Broadly \SPSE makes three major changes in KLEE: \textit{generating speculative symbolic states}, \textit{propagating potentially sensitive data} and \textit{symbolically modeling the cache behaviour}.

\paragraph{\textbf{Generating speculative symbolic states.}} 
A symbolic execution engine interprets a single instruction symbolically subject to the constraints imposed on the respective symbolic state. The initial symbolic state is constrained via the logical formula {\em true}. If the constraint imposed on the current symbolic state is $C$ and the engine encounters a branch instruction with condition $\phi_b$, then traditional symbolic execution engines check the satisfiability of constraints $C \wedge \phi_b$ and $C \wedge \neg \phi_b$. If such a constraint is satisfiable, then the engine creates a new symbolic state with the constraint. The new state inherits the state before encountering the branch instruction, but proceeds interpreting the subsequent instructions independently. Our \SPSE approach generates two extra symbolic states to model the speculative execution. These states are generated to model the speculative paths and they also model nested speculative execution. We also modify the path selection heuristic in KLEE to take into account the newly generated speculative symbolic states. Specifically, when the scheduler selects a normal state $S_m$ to execute, we check whether the state may be immediately preceded by any speculative state. If such is the case, then \SPSE selects a speculative state $SS_i$ to process. The normal state $S_m$ is not processed until all preceding speculative states of $S_m$ are handled. \SPSE can use all existing state selection strategies in KLEE, such as Depth First Search (DFS), Breadth First Search (BFS), random path selection (random-path) for both the normal state selection and speculative state selection.


\paragraph{\textbf{Propagating potentially sensitive data.}} \SPSE propagates the sensitive data along the execution path to identify the addresses that may leak the sensitive data to the cache state. When a memory load instruction reads a variable $v_s$ from an out-of-bound memory location, we mark $v_s$ as sensitive. 
All new expressions dependent on $v_s$ are subsequently marked sensitive as well. By tracking these sensitive expressions, we can detect if a memory access leads to the leakage of sensitive data. This is accomplished by checking whether the accessed memory address is constructed from any  sensitive expressions. 

\paragraph{\textbf{Symbolically modeling the cache behaviour.}} Our \SPSE tool models the cache to further check whether a cache state impacted by a sensitive address can be observed in an execution point, in particular, at the termination of a program for the access-based cache side-channel attack. Each execution state contains a cache state that symbolically records the cache content along the execution path. The cache modeling of \SPSE collects all memory load and store addresses except the memory store addresses in a speculative execution. There exists multiple reasons for such a design choice. Firstly, the memory store is not visible to the cache until the speculatively executed instructions are committed in the real execution of a processor. Secondly, our assumption is that all speculative executions in \SPSE are caused by the branch mis-prediction and all speculatively executed instructions are rolled back. Upon the termination of an execution, the symbolic cache model is constructed in line with the explanation in Section~\ref{sec:cache} and we call the STP solver to check whether the sensitive address may still stay in the cache. 


\section {Evaluation}
\label{sec:evaluation}
In this section, we perform the effectiveness evaluation of \SPSE in detecting the Bounds Check Bypass (BCB or Spectre variant 1) vulnerabilities. We aim to answer the following research questions:
\begin{enumerate}
	\item {\bf RQ1:} Can \SPSE effectively detect various kinds of BCB vulnerabilities?
	\item {\bf RQ2:} How efficient is \SPSE in detecting the BCB vulnerabilities?
	\item {\bf RQ3:} How effective is our cache model in detecting cache side-channel leakage through speculative path?
\end{enumerate}
Note that BCB vulnerability has not been reported in the wild yet. Therefore, we first run \SPSE on the litmus tests created by Kocher~\cite{spectremitigations}. These litmus tests are different 
types of Spectre vulnerable code patterns. 
Secondly, we run \SPSE on a set of security-critical benchmarks to check whether \SPSE can find the potential BCB vulnerabilities. 
Finally, we evaluate the effectiveness of our cache model in \SPSE by modifying the litmus tests and the security-critical benchmarks appropriately.

\subsection {Evaluation of \SPSE on litmus tests}
\begin{lstlisting}[float,floatplacement=H, basicstyle=\small, xleftmargin=2em, numbersep=1em, caption={The code for testing \SPSE with cache modelling. }\label{lst:litums_code}]
int array1_size = 16;
char array1[16];
char array2[256];
char array3[512 * 64];
char temp = 0;
char victim_fun(int idx) {
    register int i;
    if (idx < array1_size) {                  
        temp &= array2[array1[idx]];
    }
#define ITER N      // N = {1, ... , 512}
    for (i = 0; i < ITER; i++) {
        temp &= array3[i * 64];
    }
    return temp;
}
void main() {
    int x;
    klee_make_symbolic(&x, sizeof(x), "x");
    victim_fun(x);
}
\end{lstlisting}
No real BCB vulnerability has been reported in the wild. So we first rely on fifteen litmus test programs with Spectre vulnerability created by Kocher~\cite{spectremitigations}. We aim to check whether \SPSE can successfully detect these different variations of BCB vulnerabilities. These litmus tests were originally developed to evaluate the effectiveness of the Spectre mitigation in Microsoft C/C++ compiler. The Microsoft compiler uses static analysis to identify the vulnerable code fragments and inserts {\tt lfence} to repair the vulnerable code. Kocher reports~\cite{spectremitigations} that the Microsoft compiler can only identify two out of 15 vulnerable programs. This is because instead of using precise static analysis, Microsoft C/C++ compiler only performs a simple code pattern matching to identify code fragments related to the BCB vulnerabilities. In contrast, \SPSE can correctly detect all the BCB variants in 15 litmus tests produced by Kocher.


The programs used in litmus tests contain no memory access after the sensitive data is leaked and brought into the cache along the speculative path. As a result, our cache modeling has no impact on the detection results and all the litmus tests are correctly confirmed to contain Spectre vulnerability by \SPSE. Thus, we design additional experiments to showcase the power of cache modeling in \SPSE by introducing memory access code in the litmus test programs after the spectre vulnerability. 

For evaluating our cache model, we use a 32~KB set-associative cache with the LRU replacement policy and each cache line has 64 bytes data. We configure the cache to be 2-way, 4-way or 8-way in our experiments. We mainly consider the \textit{PRIME + PROBE} attack on L1 cache. This attack is used to target both data \cite{Osvik2006,Percival05cachemissing} and instruction cache~\cite{Onur2010}. 

A modified litmus test code is outlined in Listing~\ref{lst:litums_code}. The code contains a vulnerable function {\tt victim\_fun()} that receives an integer {\tt idx} as an argument. The {\tt if} statement at line~8 checks whether {\tt idx} is less than the {\tt array1[]} size {\ array1\_size}. If the condition holds, then {\tt idx} is used to access {\tt array1[]}. The code between line~8 and 10 exposes a typical BCB vulnerability. Specifically, if the branch at line~8 is mis-predicted, then the access of {\tt array1[]} with {\tt idx} value greater than {\tt array1\_size} can bring in potentially sensitive data. This is because {\tt array1[idx]} can point outside of {\tt array1[]} when the branch at line~8 is mis-predicted. The sensitive data can subsequently be leaked to the cache state by accessing {\tt array2[]} at line~9. The question remains whether the leaked data will still remain in the cache after completion of the program execution. This question can be answered by our cache modeling in \SPSE.

Thus, to test the effectiveness of the cache model in \SPSE, we add a loop at lines~11-15. The loop continuously brings in data to the different cache sets after the leakage of sensitive data at line~9. The memory accesses in the loop may evict the sensitive data introduced into the cache by BCB vulnerability (line~9) after {\tt N} iterations. Each iteration brings a memory block to a different cache line, for example, the loop introduces a memory block for each cache set by the first 256 iterations and the entire cache is filled up after performing total 512 iterations for a 2-way (256 sets) cache. We run this litmus test program with different value of {\tt N} from 1 to 512 to evaluate the effectiveness of \SPSE. Specifically, we aim to detect the eviction of the sensitive data from the cache for different values of {\tt N} and different cache associativities (i.e. 2, 4 and 8).  

\begin{figure} 
	\includegraphics[width=0.5\textwidth]{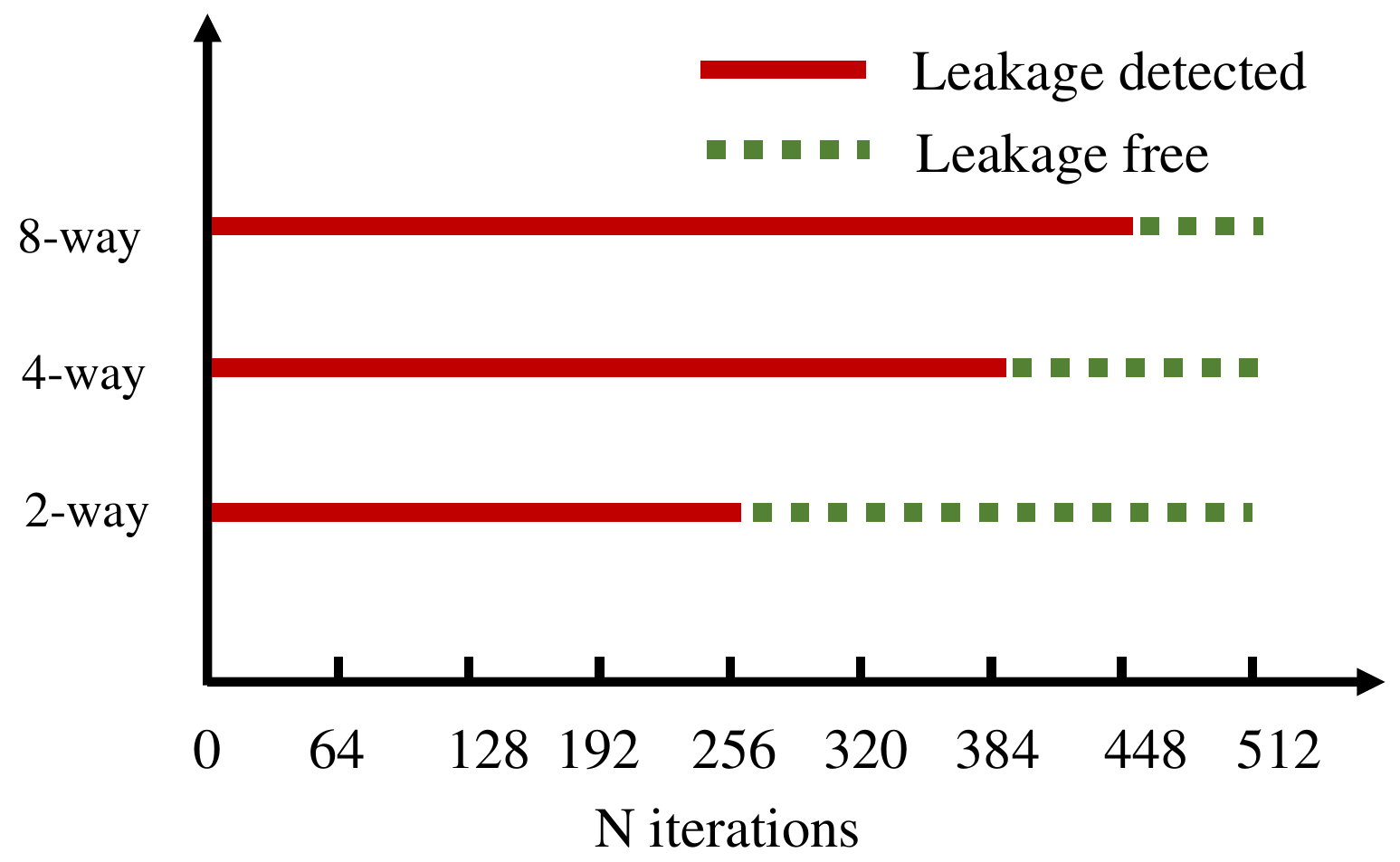}
		\caption{The detection result of \SPSE with cache model enabled. \#-way represents a cache setting with \#-way set associative cache.}
	\label{fig:litmus_result}
\end{figure}

The outcome of our findings is shown in Figure~\ref{fig:litmus_result}. The red solid line denotes that \SPSE can detect the sensitive cache state, which means the sensitive data is still in the cache after {\tt N} memory accesses in the test code. In contrast, the green dash line indicates that the sensitive data has been evicted from the cache by the additional code (Leakage free). We can see from Figure~\ref{fig:litmus_result} that the sensitive data is no longer present in the cache after 260, 288 and 452 memory accesses for cache associativity 2, 4 and 8, respectively. 

The result in Figure~\ref{fig:litmus_result} proves the effectiveness of \SPSE cache modeling. As an example, consider the code at line~9 in Listing~\ref{lst:litums_code}. The data read by {\tt array1[idx]} is one byte represented as $B_i$. Thus, $B_i$ has a value between 0 and 255. The address of the memory access performed by {\tt array2[$B_i$]} is captured via $array2+B_i$. As the least significant six bits of $B_i$ are used for the byte offset in the cache block (64 byte cache block), only two bits of $B_i$ are used for the two least significant bits of the cache set index. Thus, the address $array2+B_i$ can map to one of four selected contiguous cache sets depending on the value of $B_i$ for any cache associativity. Thus to completely evict $array2+B_i$ from the cache for arbitrary values of $B_i$, we need access to 8, 16 and 32 corresponding caches lines for 2, 4 and 8-way associate caches, respectively. 

As shown in Figure~\ref{fig:litmus_result}, for 2-way set associative caches, the leakage is undetectable after 260 memory accesses from the loop at lines~11-13. In a 2-way set-associative cache, $array2+B_i$ can potentially map to  four contiguous cache sets depending on the value of $B_i$. 
Thus, if we want to guarantee the eviction of $array2+B_i$ from the cache, then we need to fill up these contiguous four cache sets that $array2+B_i$ may map to. In our experiment, $array2$ was mapped to the first cache set. As a result, 260 memory accesses can completely fill up the first four sets of a 2-way cache. Specifically, the first 256 iterations of the loop (lines~11-13) access memory blocks mapping to all cache sets (256 cache sets  for 2-ways cache) and the rest four iterations introduce the second memory blocks for first four cache sets. This guarantees the removal of $array2+B_i$ from the cache for any value of $B_i$. 
To the best of our knowledge, none of the existing tools such as  oo7~\cite{wang2018oo7} and SPECTECTOR\cite{guarnieri2018spectector} can accurately verify the cache-side channel freedom against BCB attack like \SPSE.

\subsection{Effectiveness and Efficiency: Detection of BCB Gadgets in Real Programs.}


\begin{table}[]
	\caption{Subject benchmarks.}
	\label{tb:bm}
	\resizebox{\linewidth}{!}{
		\begin{tabular}{lllrr}
			\hline 
			\textbf{Program}           & \textbf{Source} & \textbf{Description}               & \textbf{LoC} & \textbf{\#Branch} \\ \hline \hline
			\textbf{chacha20} & LibTomCrypt     & chach20poly1305 cipher             & 776          & 71                \\ 
			\textbf{aes}      & LibTomCrypt     & AES implementation                 & 1,838        & 27                \\ 
			\textbf{encoder}  & LibTomCrypt     & encode binary data to ASCII string & 134          & 14                \\ 
			\textbf{ocb3}     & LibTomCrypt     & OCB implementation                 & 377          & 40                \\ 
			\textbf{salsa}    & Linux-tegra     & Salsa20 stream cipher              & 279          & 20                \\ 
			\textbf{camellia} & Linux-tegra     & camellia cipher                    & 1,324        & 12                \\ 
			\textbf{seed}     & Linux-tegra     & Seed cipher                        & 487          & 9                 \\ 
			\textbf{str2key}  & openssl         & Key preparation for DES            & 385          & 12                \\ 
			\textbf{des}      & openssl         & DES implementation                 & 1,051        & 11                \\ 
			\textbf{hash}     & hpn-ssh         & hash function                      & 304          & 24                \\ \hline \hline
	\end{tabular}
}
\end{table}
\textbf{Benchmark selection.} To evaluate \SPSE on real programs, we select ten cryptography related programs from well known projects: {\tt libTomCrypt}, {\tt hpn-ssh}, {\tt openssl} and {\tt Linux-tegra}. Table~\ref{tb:bm} outlines some salient features of the subject benchmarks. All the benchmarks potentially process or contain sensitive data. All of these benchmarks were also used in a recent work~\cite{wu2019abstract} to perform the analysis of speculative execution via abstract interpretation. In Table~\ref{tb:bm}, column {\em LoC} denotes the lines of code; the collected programs have 134 ({\tt encoder}) to 1,838 ({\tt AES}) lines of code. The column {\tt \#Branch} denotes the number of branches in each program ranging from 9 ({\tt seed}) to 71 ({\tt chacha20}). For all the benchmarks, we use the internal function {\tt klee\_make\_symbolic()} of 
KLEE to set the input of the programs (e.g., the plaintext and the key in cryptography programs) as symbolic variables. All benchmarks are compiled by Clang-6.0 with "-O1" optimization.

\textbf{Experimental results.} We run KLEE, \SPSE with SEW=50, and \SPSE with SEW=100 (SEW is the size of the speculative window in terms of the number of micro-instructions) on the benchmarks listed in Table~\ref{tb:bm} to compare the performance and the effectiveness of \SPSE to detect BCB gadgets. The results are shown in Table~\ref{tb:detection}. The column {\em Explored paths} denotes the number of explored normal execution paths and the column {\em Explored speculative path} indicates the explored speculative execution paths by \SPSE. 

In each category of KLEE, \SPSE 50 and \SPSE 100, column {\em Analysis time} provides the analysis time of the tool. We conduct our experimental evaluation on Intel Xeon Gold 6126~\cite{xeongold} running at 2.6GHz with 192GB memory. Intel Xeon Gold 6126 is equipped with 12 cores (24 threads) and 19.25MB shared last-level cache (LLC). The machine is running a Ubuntu 16.04 server with Linux kernel 4.4. 

Both KLEE, \SPSE 50, \SPSE 100 complete the analysis within 69 seconds. More specifically, for most benchmarks, \SPSE 50 and \SPSE 100 have longer analysis time than KLEE; but the analysis time of \SPSE is still acceptable. For example, KLEE explores three paths of {\tt chacha20} in 0.50s, \SPSE 50 explores all three normal paths along with 12,392 speculative paths in 2s. Besides, \SPSE 100 always explores more speculative paths than \SPSE 50 because \SPSE 100 executes more instructions along any speculative path.  Moreover, if \SPSE encounters branch instructions along the speculative path, then it generates new speculative states (nested speculative execution), resulting in managing larger number of symbolic states as compared to KLEE. Finally, the speculative execution might be terminated upon receiving an exception or the program exit event. The column {\em Avg. \#inst} in Table~\ref{tb:detection} shows the average number of the instructions executed along the speculative path, which is close to the SEW value in most benchmarks (e.g., 47.49 and 95.25 for \SPSE 50 and \SPSE 100, respectively, while analyzing {\tt chacha20}).   


\begin{table*}[]
	\caption {The analysis performance comparison of KLEE, \SPSE 50 and \SPSE 100 along with the detection results of BCB gadgets. Avg. \#inst= The average number of instructions executed on the speculative path. VB= vulnerable branch. UC\_VB= user controlled vulnerable branch. RS=Read Secret. LS=Leak Secret.}
	\label{tb:detection}
	\resizebox{\linewidth}{!}{
		\begin{tabular}{|l|r|r|r|r|r|r|r|r|r|r|r|r|r|r|r|r|r|r|}
			\hline
			\multirow{2}{*}{Program} & \multicolumn{2}{c|}{\textbf{KLEE}}                                                                                                     & \multicolumn{8}{c|}{\textbf{\SPSE 50}}                                                                                                                                                                                                                                                                                                                    & \multicolumn{8}{c|}{\textbf{\SPSE 100}}                                                                                                                                                                                                                                                                                                                   \\ \cline{2-19} 
			& \textbf{\begin{tabular}[c]{@{}l@{}}Analysis \\ time\end{tabular}} & \textbf{\begin{tabular}[c]{@{}l@{}}Explored \\ paths\end{tabular}} & \textbf{\begin{tabular}[c]{@{}l@{}}Analysis \\ time\end{tabular}} & \textbf{\begin{tabular}[c]{@{}l@{}}Explored \\ paths\end{tabular}} & \textbf{\begin{tabular}[c]{@{}l@{}}Explored \\ speculative \\ paths\end{tabular}} & \textbf{\begin{tabular}[c]{@{}l@{}}Avg. \\ \#inst\end{tabular}} & \textbf{VB} & \textbf{UC\_VB} & \textbf{RS} & \textbf{LS} & \textbf{\begin{tabular}[c]{@{}l@{}}Analysis \\ time\end{tabular}} & \textbf{\begin{tabular}[c]{@{}l@{}}Explored \\ paths\end{tabular}} & \textbf{\begin{tabular}[c]{@{}l@{}}Explored \\ speculative \\ paths\end{tabular}} & \textbf{\begin{tabular}[c]{@{}l@{}}Avg. \\ \#inst\end{tabular}} & \textbf{VB} & \textbf{UC\_VB} & \textbf{RS} & \textbf{LS} \\ \hline \hline
\textbf{chacha20} & 0.50s & 3   & 2s    & 3   & 12392 & 47.49 & 8 & 0 & 6 & 0 & 12s   & 3   & 124364 & 95.25 & 14 & 0 & 7 & 0 \\ \hline
\textbf{aes}      & 0.06s & 1   & 0.06s & 1   & 524   & 47.75 & 2 & 0 & 2 & 0 & 0.16s & 1   & 547    & 92.67 & 2  & 0 & 2 & 0 \\ \hline
\textbf{encoder}  & 0.45s & 22  & 4s    & 22  & 2090  & 42.64 & 2 & 1 & 3 & 0 & 11s   & 22  & 10502  & 81.55 & 2  & 1 & 3 & 0 \\ \hline
\textbf{ocb3}     & 0.11s & 2   & 0.22s & 2   & 6286  & 49.61 & 2 & 0 & 2 & 0 & 1s    & 2   & 58859  & 99.52 & 5  & 0 & 5 & 0 \\ \hline
\textbf{salsa20}  & 0.08s & 2   & 0.26s & 2   & 308   & 45.8  & 2 & 0 & 2 & 0 & 0.44s & 2   & 556    & 82.07 & 2  & 0 & 2 & 0 \\ \hline
\textbf{camellia} & 22s   & 4   & 22s   & 4   & 3141  & 44.98 & 1 & 0 & 1 & 0 & 23s   & 4   & 10440  & 82.07 & 1  & 0 & 1 & 0 \\ \hline
\textbf{seed}     & 19s   & 1   & 20s   & 1   & 242   & 49.45 & 1 & 0 & 1 & 0 & 20s   & 1   & 370    & 99.1  & 1  & 0 & 1 & 0 \\ \hline
\textbf{str2key}  & 41s   & 114 & 48s   & 114 & 2101  & 49.81 & 2 & 0 & 1 & 1 & 69s   & 114 & 8500   & 99.51 & 2  & 0 & 1 & 1 \\ \hline
\textbf{des}      & 0.01s & 1   & 0.01s & 1   & 8     & 6.88  & 1 & 0 & 1 & 0 & 0.01s & 1   & 8      & 6.88  & 1  & 0 & 1 & 0 \\ \hline
\textbf{hash}     & 0.12s & 1   & 0.16s & 1   & 1513  & 49.41 & 1 & 0 & 1 & 0 & 0.23s & 1   & 3278   & 99.44 & 1  & 0 & 1 & 0 \\ \hline
		\end{tabular}
	}
\end{table*}

As for the detection result of BCB Gadgets, the detected number of vulnerable instructions are listed in columns {\em VB, UC\_VB, RS} and {\em LS}. {\em VB} represents the number of vulnerable branches. The mis-prediction of such branches may result in the secret data to be loaded in the cache. The term {\em UC\_VB} means that the vulnerable branch can directly be trained via the user controlled input. {\em RS} is the abbreviation of Read Secret. Specifically, {\em RS} means that the secret can be loaded after executing the respective instruction. {\em LS} is an abbreviation of Leak secret wherein an instruction can leak the secret loaded by {\em RS} instruction to the cache state. The columns {\em VB, UC\_VB, RS} and {\em LS} in Table~\ref{tb:detection} are reported as the unique code locations and if one vulnerable code location appears in several speculative execution paths, the code location is only reported once.  

We detect {\em VB} and {\em RS} in all the benchmarks. For example, \SPSE 50 found eight vulnerable branches in {\tt chacha20} but none of of them is user-controlled. Only the benchmark {\tt str2key} contains a Leak secret ({\em LS}), which means that the secret can potentially be loaded to the cache and observed by the attacker. 

\begin{lstlisting}[basicstyle=\normalsize, caption={Potential Spectre variant 1 vulnerability in {\tt str2key}; TB, RS, LS are highlighted.}\label{lst:str2key}]
void DES_set_odd_parity(DES_cblock*key) {
    int i;
    for (i=0; i<DES_KEY_SZ; i++)  /* VB */
        (*key)[i]=odd_parity[(*key)[i]]; /* RS, LS */
}
\end{lstlisting}

Listing~\ref{lst:str2key} shows a potential Spectre variant 1 vulnerability  in the {\tt str2key} function {\tt DES\_set\_odd\_parity()}. The loop iteratively reads the data pointed by {\tt *key} and uses the data to index array {\tt odd\_parity}. A mis-prediction of the {\tt for} loop condition may cause a speculative execution of a few more loop iterations than normal execution.  This may lead sensitive data beyond the end of {\tt *key} (i.e. beyond the size DES\_KEY\_SZ) to be loaded into the cache. The sensitive data can impact the cache state when it is used to access array {\tt odd\_parity}. Thus, the cache state can potentially be observed by the attacker through probing array {\tt odd\_parity}. The exact amount of the leakage depends on the number of iterations that can be speculatively executed. However, in its current state, \SPSE does not compute an exact quantification of the leakage. 


We also compare the \SPSE result with oo7~\cite{wang2018oo7} and show that oo7 can only detect data leakage in {\tt encoder} and {\tt ocb3}. This is because oo7 only identifies the user controlled branches as vulnerable branches. However, \SPSE assumes all branches can be mis-trained by the attacker; for example, the victim process and the attacker process may be scheduled to the same core and the attacker can directly mis-train the branch prediction~\cite{canella2018systematic,Evtyushkin}.  

\subsection{Leakage detection with cache modeling.}
\begin{table*}[]
		\caption{The detection result with cache modeling enabled for different cache configurations.}
		\label {tb:cache}
		\resizebox{\linewidth}{!}{
	\begin{tabular}{|l|r|r|r|r||r|r|r|r|r|r|r|r|r|}
		\hline
		\multicolumn{1}{|c|}{\multirow{3}{*}{\textbf{Program}}} & \multicolumn{4}{c|}{\multirow{2}{*}{\textbf{\SPSE 100}}}                                                                                                                                                                                                                                & \multicolumn{9}{c|}{\textbf{\SPSE 100 with cache modeling}}                                                                                                                                                                                                                                                                                                                                                                                                                                                                                                                                                                                   \\ \cline{6-14} 
		\multicolumn{1}{|c|}{}                                  & \multicolumn{4}{c|}{}                                                                                                                                                                                                                                                                      & \multicolumn{3}{c|}{\textbf{2-ways}}                                                                                                                                                                           & \multicolumn{3}{c|}{\textbf{4-ways}}                                                                                                                                                                           & \multicolumn{3}{c|}{\textbf{8-ways}}                                                                                                                                                                           \\ \cline{2-14} 
		\multicolumn{1}{|c|}{}                                  & \textbf{\begin{tabular}[c]{@{}l@{}}Symbolic \\ address (\%)\end{tabular}} & \textbf{\begin{tabular}[c]{@{}l@{}}Analysis\\  time\end{tabular}} & \textbf{\begin{tabular}[c]{@{}l@{}}Detected \\ leakage\end{tabular}} & \textbf{\begin{tabular}[c]{@{}l@{}}Solver \\ time(\%)\end{tabular}} & \textbf{\begin{tabular}[c]{@{}l@{}}Analysis \\ time\end{tabular}} & \textbf{\begin{tabular}[c]{@{}l@{}}Detected \\ leakage\end{tabular}} & \textbf{\begin{tabular}[c]{@{}l@{}}Solver \\ time(\%)\end{tabular}} & \textbf{\begin{tabular}[c]{@{}l@{}}Analysis \\ time\end{tabular}} & \textbf{\begin{tabular}[c]{@{}l@{}}Detected \\ leakage\end{tabular}} & \textbf{\begin{tabular}[c]{@{}l@{}}Solver \\ time(\%)\end{tabular}} & \textbf{\begin{tabular}[c]{@{}l@{}}Analysis \\ time\end{tabular}} & \textbf{\begin{tabular}[c]{@{}l@{}}Detected \\ leakage\end{tabular}} & \textbf{\begin{tabular}[c]{@{}l@{}}Solver \\ time(\%)\end{tabular}} \\ \hline \hline
chacha20 & 5.56\%  & 131s  & 3 & 56.09\% & 1256s & 3 & 3.28\%  & 1217s & 3 & 3.30\%  & 1288s & 3 & 3.18\%  \\ \hline
aes      & 0.52\%  & 0.30s & 3 & 45.43\% & 62s   & 3 & 12.86\% & 10s   & 1 & 4.53\%  & 10s   & 1 & 10.00\% \\ \hline
encoder  & 27.72\% & 3s    & 3 & 59.96\% & 6s    & 2 & 42.01\% & 5s    & 2 & 43.35\% & 5s    & 2 & 43.87\% \\ \hline
ocb3     & 2.49\%  & 5s    & 3 & 22.19\% & 11s   & 2 & 12.80\% & 11s   & 2 & 13.12\% & 10s   & 2 & 13.84\% \\ \hline
salsa20  & 8.52\%  & 1s    & 3 & 10.51\% & 1s    & 2 & 8.90\%  & 1s    & 2 & 8.50\%  & 1s    & 2 & 7.62\%  \\ \hline
camellia & 13.53\% & 19s   & 3 & 86.98\% & 1712s & 1 & 1.27\%  & 1696s & 1 & 1.32\%  & 1670s & 1 & 1.26\%  \\ \hline
seed     & 25.90\% & 24s   & 3 & 92.13\% & 1074s & 3 & 10.87\% & 1174s & 2 & 8.71\%  & 1036s & 3 & 10.87\% \\ \hline
str2key  & 12.12\% & 793s  & 4 & 64.90\% & 1.08h & 4 & 43.23\% & 0.89h & 4 & 31.38\% & 0.82h & 4 & 29.05\% \\ \hline
des      & 25.00\% & 59s   & 3 & 88.76\% & 37s   & 3 & 77.54\% & 72s   & 3 & 86.90\% & 46s   & 3 & 79.59\% \\ \hline
hash     & 3.76\%  & 19s   & 3 & 87.74\% & 317s  & 2 & 4.73\%  & 120s  & 3 & 7.38\%  & 318s  & 2 & 4.72\%  \\ \hline
	\end{tabular}
}
\end{table*}

The cache modeling of \SPSE can accurately check whether the leaked sensitive data can be observed by the attacker through cache side-channel attack. Our cache model is not invoked until some sensitive data leakage 
is identified along the speculative path. As we do not find any data leakage in our benchmarks, in this experiment, we insert several vulnerable functions to the benchmarks and check whether \SPSE can detect them. More specifically, we randomly choose three Spectre v1 variant functions suggested by Kocher~\cite{spectremitigations}, then insert them to the start, middle and the end of each benchmark listed in Table~\ref{tb:bm}. Then, we run \SPSE with cache modeling enabled. Each experiment is conducted over three runs for three different cache associativities: 2, 4 and 8.  

Table~\ref{tb:cache} shows the test result of comparing \SPSE 100 and \SPSE 100 with cache model enabled (\SPSE100\_Cache). All inserted vulnerable code that leaks the sensitive data in the speculative execution path have been detected by \SPSE 100 ({\tt str2key} contains one original leakage). More importantly, we observe that the number of vulnerable code fragments reduces when we enable the cache modeling. For example, three data leakage scenarios were identified in {\tt ocb3} without cache modeling; but only two of them were identified when cache modeling is enabled. The remaining two data leakages were identified as false positives. This means that the sensitive data loaded into the cache were subsequently evicted by other memory accesses. Moreover, we observe that the presence of information leakage might depend on the cache configuration, asserting further importance to the cache modeling embodied by \SPSE. For example, three data leakage scenarios were detected by \SPSE for 2-ways cache in {\tt aes}; but only two leakage scenarios were detected when using the 4-ways or 8-ways cache configurations for the same {\tt aes} benchmark. 

The precision of \SPSE comes with the cost of solving the symbolic cache model. Thus both the analysis time and the solver time increase (as compared to the cache modeling being disabled). The time to solve the symbolic cache model  depends on the number of memory accesses and the percentage of symbolic addresses. As observed from Table~\ref{tb:cache} that the percentage of symbolic addresses is relatively low (the maximum is 27.72\% for {\tt encoder}); thus the solver can finish within an acceptable time. Finally, except for {\tt hash} and {\tt des}, we did not observe a noticeable difference in analysis time with increased cache associativity. This means that our symbolic cache model scales well with respect to varying cache configurations. 

\section{Threats to validity}
\label{sec:threats}
\textbf{Path explosion.} Path explosion is a major challenge in the symbolic execution. \SPSE is based on symbolic execution, which does not scale to large programs while exploring all feasible program paths. In particular, \SPSE forks more paths than the classical symbolic execution for performing the speculative execution. However, only limited number of instructions are executed on the speculative paths in \SPSE, which is bounded by the Speculative Execution Windows (SEW). Thus, as observed in our evaluation,  \SPSE has similar complexity with the classical symbolic execution. Moreover, the existing methods to alleviate the path explosion can also be used by \SPSE, for example, state merging~\cite{hansen2009state, kuznetsov2012efficient}. Specifically, the speculative states can be merged similarly as the classical states when the control flows of a program merges. Besides, the symbolic execution can be guided by the low-cost static analysis in such a fashion that a static analysis can be performed to roughly locate the vulnerable code and prune the redundant paths during the construction of symbolic execution tree. 

\textbf{Precise modeling of program behavior.} The program behavior running on the hardware may not be the same as it is in the symbolic execution. This is because \SPSE uses bitcode that may not replicate exactly the same behavior as the final binary code due to the compiler optimization. For example, the program may have more memory accesses when running on the hardware than it is during the symbolic execution due to the register spilling. However, \SPSE is designed as an over-approximation method that it captures all necessary memory accesses and detects all potentially secret leakage. This results in the absence of false negatives. In other words, \SPSE guarantees that all leakage in the real execution can be detected. However, \SPSE can generate false positives. For instance, the leakage detected by \SPSE may not be exploitable in the real hardware. 
\section {Related work}
\label{sec:related_work}

\textbf{Spectre-style attack mitigation. } Several approaches have been proposed to mitigate Spectre vulnerabilities~\cite{slharden, oleksii, developerguidance, wang2018oo7,guarnieri2018spectector}.   

Speculative Load hardening~\cite{slharden} (SLH) is a mitigation technique for Spectre variant 1, adopted by the LLVM compiler. SLH identifies the potentially vulnerable code fragments where memory accesses depend on the conditional branches and then inserts hardening instruction sequence to nullify the pointers that may leak the data. 
SLH hardens the \textbf{RS} stage of the vulnerable code, the secret data cannot be loaded to the cache during speculative execution after nullifying the crucial pointers. As SLH repairs the program at every conditional branch and the hardening instructions slow down execution, it introduces 36\% performance overhead. Oleksenko et al.~\cite{oleksii} present mitigation of Spectre variant 1 attack by delaying the execution of the vulnerable instructions via introduction of artificial data dependencies instead of serialization instructions to stop speculative execution altogether. These methods lack accurate analysis and hence overestimate vulnerable code fragments leading to a significant performance overhead of the repaired programs. 

Microsoft Visual C/C++ compiler~\cite{developerguidance} enables mitigation of Spectre Variant 1 through a compiler option that inserts "lfence" serializing instruction at potentially vulnerable code. However, this technique successfully mitigates only 2 out of 15 Spectre litmus tests~\cite{Kocher2018spectre}.

oo7~\cite{wang2018oo7} is the first work proposed to mitigate Spectre-style attacks via modeling speculative execution in static analysis. oo7 works on binary and leverages taint analysis to track the vulnerable branches and memory operations that lead to Spectre-style vulnerabilities. oo7 can effectively detect and fix Spectre-style attacks, but may still produce false positives due to conservative static analysis.  

SPECTECTOR~\cite{guarnieri2018spectector} presents a principled approach using speculative non-interference in symbolic execution to discover data leakage. However, Spectector only finds whether some secret data has been speculatively accessed; it does not check the possibility of follow-up cache side-channel attack, which is what we achieved via our cache modeling.


\textbf{Side-channel attack identification via cache modeling.}
Casym~\cite{brotzman2019casym} presents a cache-aware symbolic execution to identify and fix cache side-channel vulnerabilities. Casym provides two cache models: the {\em infinite cache model} of caches with infinite size and associativity, and the {\em age model} that tracks the distance of a memory access from its most recent access. However, the description of the cache models in Casym are sketchy and hence prevents reproducibility. Besides, Casym does not consider speculative execution in its models. CACHEFIX~\cite{chattopadhyay2018symbolic} is another cache side-channel verification tool that can detect and fix the attack via symbolic execution. It also targets cache timing attacks and does not consider speculative execution paths. 

Abstract interpretation is a static analysis approach that has been effectively adopted for cache hit/miss modeling in Worst-Case Execution Time (WCET) estimation. Wu et al.~\cite{wu2019abstract} introduce abstract interpretation to side-channel attack detection by extending it to cover speculative execution. This approach targets timing based side-channel attack but does not handle Spectre attack. A similar approach is embodied  CacheAudit~\cite{doychev2015cacheaudit}, however, the CacheAudit approach does not consider speculative execution semantics. 

In summary, the previous works either do not model speculative execution or lack a precise cache model. \SPSE is the first work to integrate speculative symbolic execution with cache modeling.

\section {Conclusion}
\label{sec:conclusion}
We have presented a new software testing tool named as \SPSE to expose the micro-architectural features to the software testing. The micro-architectural features such as speculative execution and caches are generally ignored by traditional software testing. This hides the vulnerabilities caused by invisible micro-architectural behaviours when a program runs on the hardware. \SPSE makes these behaviours visible in the software testing via modeling the speculative execution and caches within the traditional symbolic execution. The experiment shows that \SPSE can effectively detect the BCB vulnerabilities and the cache model can make such detection more accurate. \SPSE is only a step forward to extend the foundation of software testing to systematically discover vulnerabilities dependent on micro-architectural features. Our tool also provides an open platform to extend our methodologies as more Spectre style attacks are being discovered. For reproducibility and further research in the area, our tool and the benchmarks are publicly available:
\begin{center}
            \url{https://github.com/winter2020/kleespectre}
\end{center}

\section*{Acknowledgments}
This research is supported by the National Research Foundation,
Prime Minister's Office, Singapore under its National Cybersecurity
R\&D Program (Award No. NRF2014NCR-NCR001-21) and administered by the National Cybersecurity R\&D Directorate.
\bibliographystyle{ACM-Reference-Format}
\balance
\bibliography{ref/references} 

\end{document}